\documentclass[letterpaper,twocolumn,prl,aps,superscriptaddress,amsmath,amssymb,floatfix]{revtex4-1}
\usepackage{mathptmx}
\usepackage[latin9]{inputenc}
\usepackage{color}
\usepackage{amsmath}
\usepackage{amssymb}
\usepackage{graphicx}
\usepackage{esint}
\usepackage{enumitem}
\usepackage[unicode=true,
 bookmarks=true,bookmarksnumbered=false,bookmarksopen=false,
 breaklinks=false,pdfborder={0 0 1},backref=false,colorlinks=true]
 {hyperref}
\hypersetup{
 linkcolor=magenta,urlcolor=blue,citecolor=blue,pdfstartview={FitH}}

\makeatletter

\pdfpageheight\paperheight
\pdfpagewidth\paperwidth

\usepackage{textcomp}
\usepackage{epstopdf}

\pdfpageheight\paperheight
\pdfpagewidth\paperwidth

\@ifundefined{textcolor}{}{%
 \definecolor{BLACK}{gray}{0}
 \definecolor{WHITE}{gray}{1}
 \definecolor{RED}{rgb}{1,0,0}
 \definecolor{GREEN}{rgb}{0,1,0}
 \definecolor{BLUE}{rgb}{0,0,1}
 \definecolor{CYAN}{cmyk}{1,0,0,0}
 \definecolor{MAGENTA}{cmyk}{0,1,0,0}
 \definecolor{YELLOW}{cmyk}{0,0,1,0}
}

\usepackage{xcolor}
\usepackage{soul}
\definecolor{blue}{rgb}{0,0,1}
\definecolor{red}{rgb}{1,0,0}
\definecolor{green}{rgb}{0,1,0}

\usepackage{soul}

\newcommand{\RNum}[1]{\uppercase\expandafter{\romannumeral #1\relax}}

\usepackage{lineno}
\modulolinenumbers[5] 

\makeatother

\begin{document}


\affiliation{Laboratory of Quantum Information, University of Science and
Technology of China, Hefei 230026, P. R. China.}
\affiliation{Center for Quantum Information, Institute for Interdisciplinary Information
Sciences, Tsinghua University, Beijing 100084, China}
\address{School of Civil Engineering, Hefei University of Technology,
Hefei, Anhui 230009, P.R. China.}
\affiliation{CAS Center For Excellence in Quantum Information and Quantum Physics,
University of Science and Technology of China, Hefei, Anhui 230026,
P. R. China.}
\affiliation{Hefei National Laboratory, Hefei 230088, China}

\title{Large-scale programmable phononic integrated circuits}

\author{Xin-Biao~Xu}
\thanks{These authors contributed equally to this work.}
\affiliation{Laboratory of Quantum Information, University of Science and Technology of China, Hefei 230026, P. R. China.}
\affiliation{CAS Center For Excellence in Quantum Information and Quantum Physics, University of Science and Technology of China, Hefei, Anhui 230026, P. R. China.}

\author{Yu~Zeng}
\thanks{These authors contributed equally to this work.}
\affiliation{Laboratory of Quantum Information, University of Science and
Technology of China, Hefei 230026, P. R. China.}
\affiliation{CAS Center For Excellence in Quantum Information and Quantum Physics,
University of Science and Technology of China, Hefei, Anhui 230026,
P. R. China.}

\author{Jia-Qi~Wang}
\affiliation{Laboratory of Quantum Information, University of Science and
Technology of China, Hefei 230026, P. R. China.}
\affiliation{CAS Center For Excellence in Quantum Information and Quantum Physics,
University of Science and Technology of China, Hefei, Anhui 230026,
P. R. China.}

\author{Zheng-Hui~Tian}
\affiliation{Laboratory of Quantum Information, University of Science and
Technology of China, Hefei 230026, P. R. China.}
\affiliation{CAS Center For Excellence in Quantum Information and Quantum Physics,
University of Science and Technology of China, Hefei, Anhui 230026,
P. R. China.}

\author{Ji-Zhe~Zhang}
\affiliation{Laboratory of Quantum Information, University of Science and
Technology of China, Hefei 230026, P. R. China.}
\affiliation{CAS Center For Excellence in Quantum Information and Quantum Physics,
University of Science and Technology of China, Hefei, Anhui 230026,
P. R. China.}

\author{Yuan-Hao~Yang}
\affiliation{Laboratory of Quantum Information, University of Science and
Technology of China, Hefei 230026, P. R. China.}
\affiliation{CAS Center For Excellence in Quantum Information and Quantum Physics,
University of Science and Technology of China, Hefei, Anhui 230026,
P. R. China.}

\author{Zheng-Xu~Zhu}
\affiliation{Laboratory of Quantum Information, University of Science and
Technology of China, Hefei 230026, P. R. China.}
\affiliation{CAS Center For Excellence in Quantum Information and Quantum Physics,
University of Science and Technology of China, Hefei, Anhui 230026,
P. R. China.}

\author{Jia-Hua~Zou}
\affiliation{Laboratory of Quantum Information, University of Science and
Technology of China, Hefei 230026, P. R. China.}
\affiliation{CAS Center For Excellence in Quantum Information and Quantum Physics,
University of Science and Technology of China, Hefei, Anhui 230026,
P. R. China.}

\author{Liantao~Xiao}
\affiliation{Center for Quantum Information, Institute for Interdisciplinary Information
Sciences, Tsinghua University, Beijing 100084, China}

\author{Weiting~Wang}
\email{wangwt2020@tsinghua.edu.cn}
\affiliation{Center for Quantum Information, Institute for Interdisciplinary Information
Sciences, Tsinghua University, Beijing 100084, China}

\author{Bao-Zhen Wang}
\address{School of Civil Engineering, Hefei University of Technology,
Hefei, Anhui 230009, P.R. China.}

\author{Guang-Can~Guo}
\affiliation{Laboratory of Quantum Information, University of Science and
Technology of China, Hefei 230026, P. R. China.}
\affiliation{CAS Center For Excellence in Quantum Information and Quantum Physics,
University of Science and Technology of China, Hefei, Anhui 230026,
P. R. China.}
\affiliation{Hefei National Laboratory, Hefei 230088, China}

\author{Luyan~Sun}
\email{luyansun@tsinghua.edu.cn}
\affiliation{Center for Quantum Information, Institute for Interdisciplinary Information Sciences, Tsinghua University, Beijing 100084, China}
\affiliation{Hefei National Laboratory, Hefei 230088, China}

\author{Chang-Ling~Zou}
\email{clzou321@ustc.edu.cn}
\affiliation{Laboratory of Quantum Information, University of Science and
Technology of China, Hefei 230026, P. R. China.}
\affiliation{CAS Center For Excellence in Quantum Information and Quantum Physics,
University of Science and Technology of China, Hefei, Anhui 230026,
P. R. China.}
\affiliation{Hefei National Laboratory, Hefei 230088, China}

\date{\today}

\begin{abstract}
\textbf{Electronic and photonic chips revolutionized information technology through massive integration of functional elements, yet phonons as fundamental information carriers in solids remain underestimated. Here, we demonstrate large-scale programmable phononic integrated circuits (PnICs) for complex signal processing. We developed a comprehensive library of gigahertz-frequency phononic building blocks that control acoustic wave propagation, polarization, and dispersion. Combining these elements, we demonstrate an ultra-compact 1$\times$128 on-chip acoustic power splitter with unprecedented integration density of 3,000/cm$^2$, a 21-port acoustic frequency demultiplexer with 3.8~MHz resolution, and a four-channel reconfigurable frequency synthesizer. This work establishes scalable phononic integration as the third pillar of information processing alongside electronics and photonics, enabling hybrid chips that combine all three domains for advanced signal processing and quantum information applications.}
\end{abstract}

\maketitle


\noindent Acoustic waves have historically served as the primary information carriers in nature, enabling communication~\cite{Chen2020,Hays1976}, navigation~\cite{Kurosawa1998,Fu2019}, and environmental sensing~\cite{Chen2014,Zhao2024,Wang2025}. In the last century, radio-frequency electromagnetic waves emerged as dominant information carriers due to their higher propagation speeds and longer transmission distances~\cite{Jornet2023,Andrews2001,Zhang2021,Koenig2013,Blais2021}. Subsequently, the development of optical fibers established photons as ideal carriers for long-distance communication, bringing the revolutionary worldwide internet~\cite{Corcoran2020,Poletti2013,Petrovich2025,Wang2022qkd}. The extraordinary advancement of both electronic (microwave) and photonic technologies has been driven largely by integrated circuit architectures, enabling massive integration of functional elements on single chips~\cite{Jayachandran2024,Bogaerts2020,Xu2024,AghaeeRad2025}. This integration has dramatically enhanced information processing capabilities while reducing costs, power consumption, and physical footprints, fueling the prosperity of modern information technology~\cite{Cao2023,Hua2025,Daudlin2025}.

Despite phonons constituting one of the three fundamental information carriers alongside electrons and photons, the application of acoustic wave in solid has remained strikingly constrained, primarily limited to isolated components such as filters and delay lines in communication systems~\cite{morgan2010surface,hashimoto2009rf,Weigel2002}. Although these applications effectively leverage the intrinsic advantages of acoustic waves, including inherently low propagation losses~\cite{Beccari2022,MacCabe2020,Xi2025}, compact wavelengths~\cite{Xu2022apl}, and strong interactions with other excitations in solids~\cite{Hackett2024,Yang2024a}, they represent only a fraction of their potential utility in complex signal processing. This limitation stems not from fundamental physical constraints but rather from the absence of a comprehensive architectural framework that would enable acoustic technologies to achieve system-level integration and information processing capabilities comparable to those realized in electronic and photonic domains.

Very recently, breakthrough has been made in the development of suspension-free phononic devices on solid-state substrates. It leverages acoustic index contrast between materials and shows remarkable potential to construct complex phononic integrated circuits (PnICs) with improved versatility~\cite{Wang2020,Mayor2021,Shao2022ne,Bicer2022,Xu2025}. Moreover, phonons as coherent information carriers have been successfully exploited in the quantum domain~\cite{Gustafsson2014,Chu2017,Satzinger2018,Qiao2023}, where they effectively store quantum information and mediate interactions between various types of qubits~\cite{Schuetz2015,Clerk2020,Wei2018}. Despite these promising developments, most previous demonstrations have been limited to the validation of individual devices with simple configurations~\cite{Xu2022apl,Wang2022}. Key questions regarding the versatility, scalability, and system-level integration of phononic circuits remain largely unexplored.

\begin{figure*}[t]
\includegraphics[width=1\textwidth]{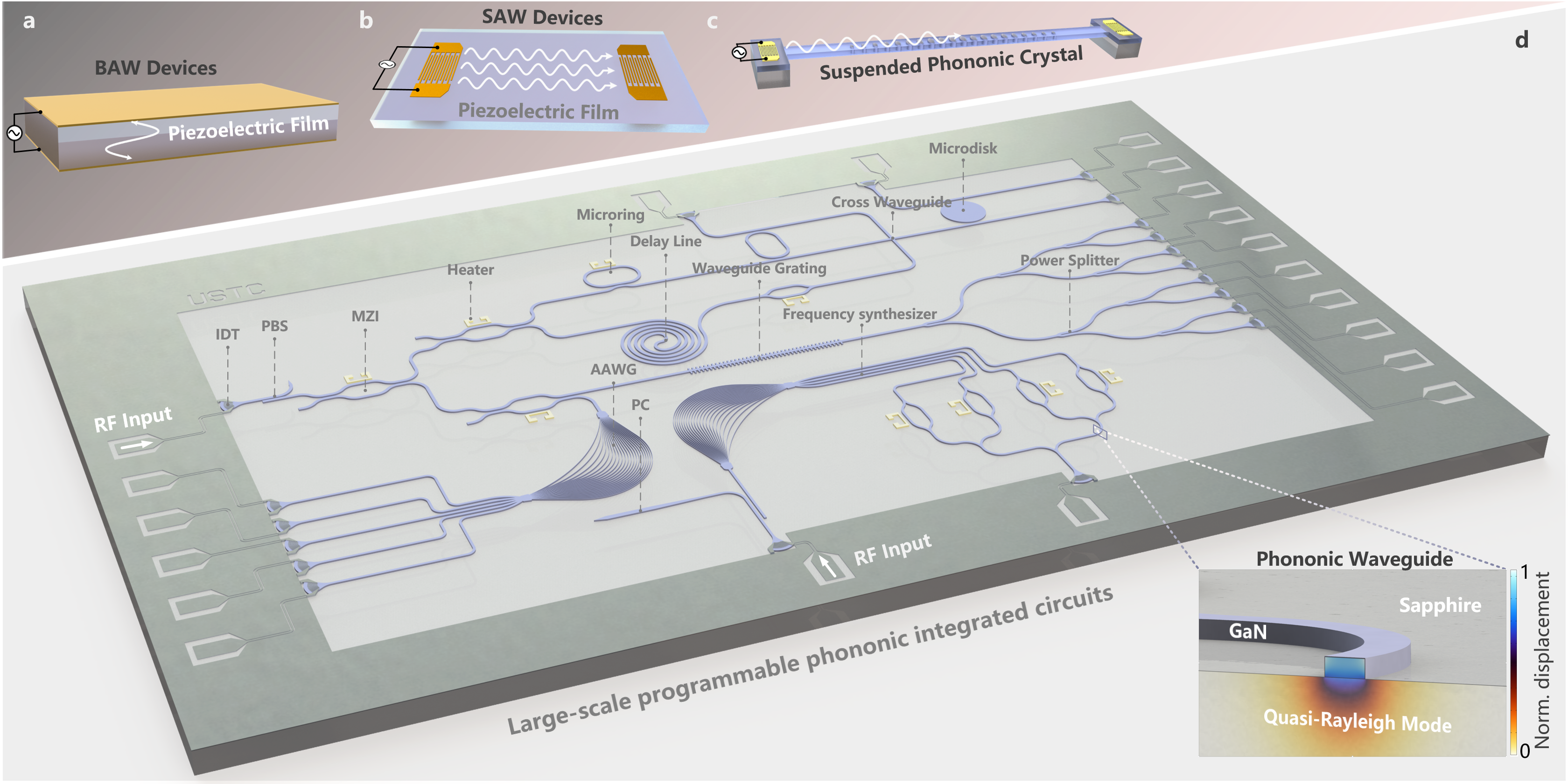}
\caption{\textbf{Large-scale programmable phononic circuits.}
\textbf{a}-\textbf{c}, Schematics of conventional integrated acoustic devices: bulk acoustic wave (BAW) (\textbf{a}), surface acoustic wave (SAW) (\textbf{b}), and suspended phononic crystal structures (\textbf{c}). \textbf{d}, Schematic of large-scale programmable suspension-free phononic integrated circuits (PnICs) leveraging phononic refractive index contrast. The inset illustrates the confined phononic mode in the phononic waveguide. Through the combination of various phononic functional building blocks, including directional couplers, Y-splitters, polarization converters (PC),  Mach-Zehnder interferometers (MZI), microring resonators, and phononic waveguide gratings, PnICs can achieve diverse phononic signal processing functionalities, such as multi-port acoustic power splitters, acoustic arrayed waveguide gratings (AAWG), and reconfigurable acoustic frequency synthesizer.}
\label{Fig1}
\end{figure*}

Here, we demonstrate large-scale programmable PnICs on a suspension-free platform, elevating phononics to system-level integration capabilities comparable to those of electronics and photonics for the first time. We first develop the key building blocks for PnICs, including phononic multi-mode interferometers, polarization rotators, phononic waveguide gratings. By combining these elements, we demonstrate an ultra-compact 1$\times$128 on-chip acoustic power splitter and an acoustic arrayed waveguide grating (AAWG), showcasing the high integration density and dispersion engineering advantages of PnICs in acoustic signal processing. Furthermore, by incorporating the AAWG, thermo-acoustic MZIs array, and power combiner/splitter, we demonstrate a reconfigurable acoustic frequency synthesizer. These achievements highlight the flexibility and programmability of PnICs for complex gigahertz acoustic and microwave signal processing tasks.

\begin{figure*}[t]
\includegraphics[width=1\textwidth]{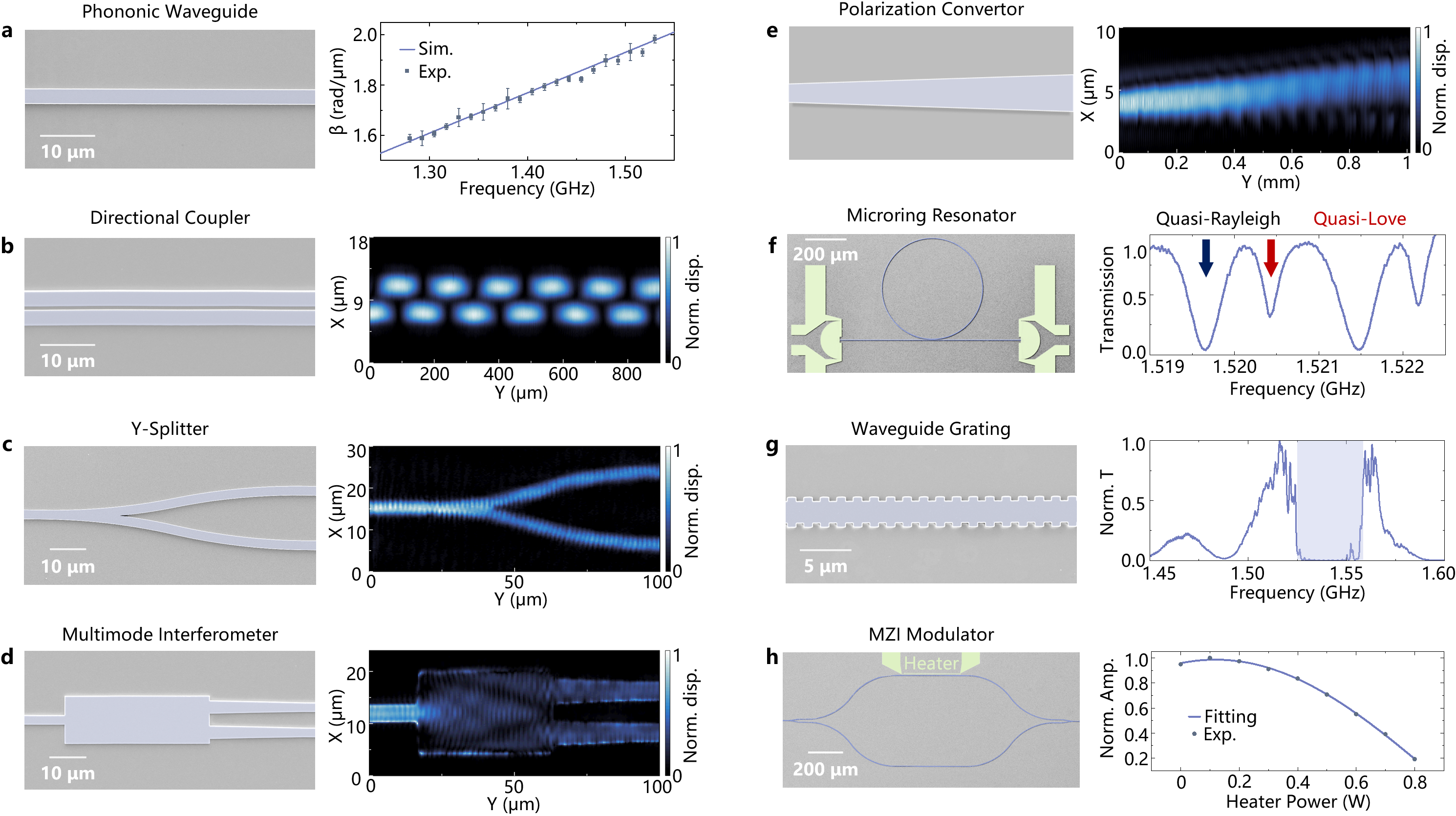}
\caption{\textbf{Building blocks of the programmable PnICs.} \textbf{a}, Scanning electron microscope (SEM) image of a dispersive phononic waveguide showing a group velocity of 3904 m/s. \textbf{b}-\textbf{d}, SEM images and the measured acoustic fields of a phononic waveguide directional coupler by home-built vibrometer (\textbf{b}), a Y-splitter (\textbf{c}), and a multi-mode interferometer for power splitting (\textbf{d}). \textbf{e},   An adiabatic polarization converter based on tapered waveguides.  \textbf{f}, SEM image and transmission spectrum of a microring resonator showing resonances of two polarization modes (quasi-Love and quasi-Rayleigh).  \textbf{g}, SEM image and transmission spectrum of a phononic waveguide grating structure with a 33~MHz bandgap. \textbf{h}, SEM image and the normalized output acoustic amplitude as a function of heater power for an MZI modulator. }
\label{Fig2}
\end{figure*}

\begin{figure*}[t]
\includegraphics[width=1\textwidth]{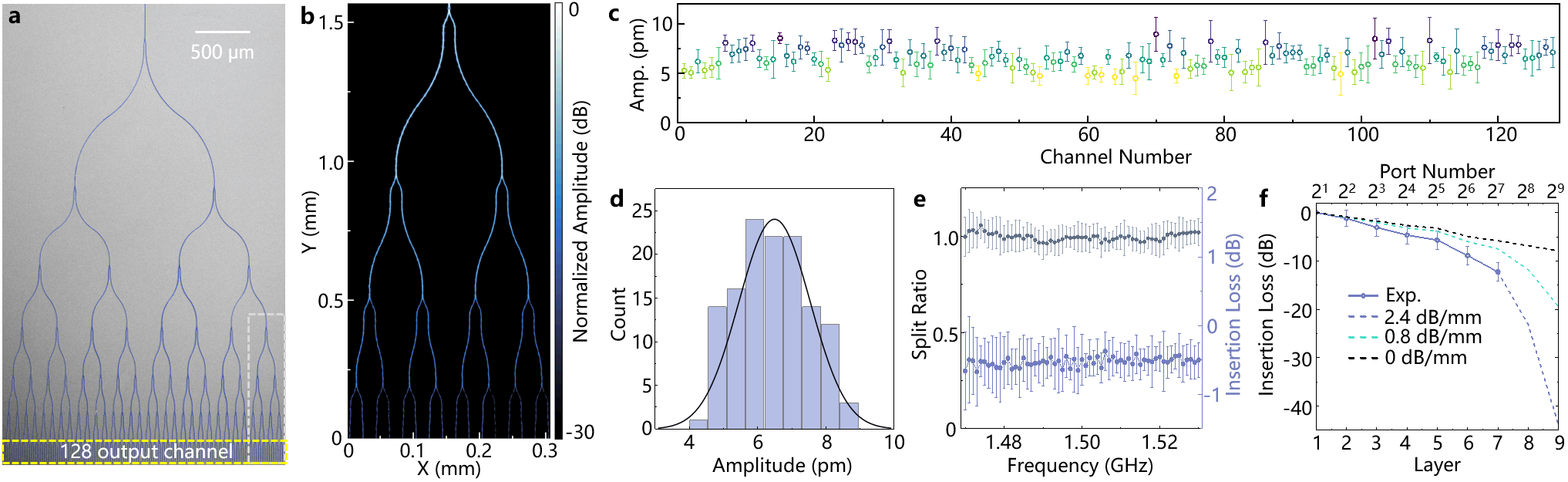}
\caption{\textbf{Phononic power splitter.} \textbf{a}, SEM image of a 1$\times$128 phononic power splitter based on a cascaded network of Y-splitters. \textbf{b}, Measured normalized acoustic amplitude distribution in the white boxed region of \textbf{a}. \textbf{c}, Measured average acoustic amplitude of the 128 output ports in the yellow boxed region of \textbf{a}. \textbf{d}, Statistical analysis of the amplitude of the 128 output ports. Black curve: normal distribution fit with mean amplitude of 6.5~pm and standard deviation of 1~pm. \textbf{e}, Split ratio and insertion loss of the power splitter as a function of frequency. \textbf{f}, Insertion loss as a function of the layer $l$ with port number $2^l$ for different waveguide losses. The dashed lines indicate the result of extrapolations.}
\label{Fig3}
\end{figure*}

\smallskip{}
\noindent \textbf{\large{}Scalable phononic integrated circuits}{\large\par}

\noindent Figures~\ref{Fig1}a-c illustrate the conventional architecture of the integrated acoustic signal processing devices. These device architectures fundamentally restrict their functionality and scalability. Bulk acoustic wave (BAW, Fig.~\ref{Fig1}a) devices utilize localized phononic modes in membrane structures, offering high quality factor resonances for narrow-band filters, but inherently lacking wave propagation capabilities required for complex signal routing~\cite{Liu2020}. Surface acoustic wave (SAW, Fig.~\ref{Fig1} b) devices are advantageous for signal delay due to the slow acoustic velocity, where a mm-long propagation length leads to microsecond-scale delays~\cite{Cho2022}. Although SAW devices utilize phonons as propagating carriers, they suffer from poor lateral confinement. This prevents efficient acoustic wave routing and the implementation of multi-port functional devices. Recent advances in suspended phononic crystal (Fig.~\ref{Fig1}c) structures have demonstrated improved capability for guiding and tightly confining acoustic waves. However, their practical implementation in large-scale circuits remains challenging due to fabrication complexities, weak mechanical stability, and low thermal dissipation rates in suspended devices~\cite{Zivari2022,Bozkurt2023}.

In contrast, we employ suspension-free PnICs to overcome these limitations, as shown in Fig.~\ref{Fig1}d. As illustrated in the inset, GHz phononic modes can be confined and propagated within compact waveguides of sub-micromete scale leveraging the acoustic refractive index contrast between different materials (gallium nitride on sapphire in this work)~\cite{Xu2025}. This phononic waveguide architecture enables the construction of various functional building blocks on PnICs, including acoustic directional couplers, power splitters, mode converters, resonators, and phase/intensity modulators. These components enable the manipulation of the path, mode, frequency, phase, and amplitude of acoustic waves. By combining these building blocks, large-scale programmable PnICs can be implemented to achieve diverse complex acoustic signal processing, such as multi-port coupling, acoustic frequency division multiplexing, and spectrum manipulation.

\smallskip{}
\noindent \textbf{\large{}Basic building blocks}{\large\par}

\noindent Figure~\ref{Fig2} presents the key building blocks we realized for our PnICs platform, enabling comprehensive control over acoustic waves across four key degrees of freedom: path, mode, frequency, and phase. In this work, we typically employ the fundamental quasi-Rayleigh mode in 700~nm thick gallium nitride waveguides, operating at a frequency around $1.5\,\mathrm{GHz}$. The devices are characterized by directly observing the acoustic vibration field distribution through a home-built scanning vibrometer system~\cite{Xu2025}. As shown in Fig.~\ref{Fig2}a, the most fundamental component is a phononic waveguide. For a waveguide with a width of $2.8\,\mathrm{\mu m}$, both experiments and numerical simulations reveal significant dispersion with a group velocity of approximately 3904~m/s that differs substantially from the phase velocity of 4879~m/s.

For spatial path control, we demonstrate three distinct power splitting approaches in Figs.~\ref{Fig2}b-d. The directional coupler (Fig.~\ref{Fig2}b) can exhibit a periodic acoustic energy distribution between adjacent waveguides through evanescent coupling, where the coupling ratio can be precisely controlled by the waveguide gap and interaction length. The measured acoustic field distribution reveals the periodic power transfer between waveguides with a coupling length of $79.14\,\mathrm{\mu m}$, in excellent agreement with theoretical predictions. In contrast, the Y-splitter (Fig.~\ref{Fig2}c) provides a broadband and symmetric 50:50 power split through direct structural bifurcation, offering a simple and robust solution. As demonstrated in our subsequent arrayed waveguide gratings, the Y-splitter holds great potential for scaling to higher port counts. Finally, power splitting can also be achieved in a multimode interferometer (Fig.~\ref{Fig2}d) through controlled modal interference by designing the structure parameters of the input ports and the multimode waveguide region.

For polarization control, we exploit the dual-polarization nature of phononic waveguides, which support both quasi-Rayleigh ($R$) and quasi-Love ($L$) modes characterized by predominantly out-of-plane and in-plane displacements, respectively. In Fig.~\ref{Fig2}e, by exploiting the mode avoided crossing in the dispersion curves for $R$ and $L$ modes and adiabatically varying the waveguide width at X-crystal orientation of sapphire, we demonstrate an efficient polarization converter~\cite{Wang2022}. Frequency domain manipulation can be achieved through resonant structures or phononic energy band engineering~\cite{Xu2022apl,Mayor2021,Bicer2022}. In Fig.~\ref{Fig2}f we demonstrate a traveling-wave microring resonator  with $R$ and $L$ two mode families. Notably, based on the distinct resonance conditions of the $R$ and $L$ modes, the microring can also function as a polarization beam splitter. It can selectively route input fields with different polarization states to different output ports. Alternatively, by introducing a periodic grating waveguide, the phononic frequency bandgaps can be engineered (Fig.~\ref{Fig2}g). By optimizing the grating period, modulation depth, and number of periods, precise spectral filtering is demonstrated in a compact $190\,\mathrm{\mu m}$ long device, exhibiting a measured bandgap of $33\,\mathrm{MHz}$.

Dynamic tuning of phononic devices can be achieved using thermoacoustic phase modulators~\cite{Shao2022,Xu2025}, which integrate thermal heaters adjacent to acoustic waveguides to effectively modulate the acoustic phase through the thermoacoustic effect. As shown in Fig.\ref{Fig2}h, we demonstrate a thermoacoustic MZI modulator. Its measured output amplitude response reveals a linear phase shift response as a function of the applied electrical power, achieving a tuning efficiency of $4.03\,\mathrm{rad/W}$ for a phase shifter length of 100~$\mu$m. All of these building blocks constitute a complete toolkit for constructing complex programmable PnICs, enabling unprecedented control over acoustic waves across all relevant degrees of freedom, as demonstrated by our large-scale integrated devices in subsequent sections.

\begin{figure*}[t]
\includegraphics[width=1\textwidth]{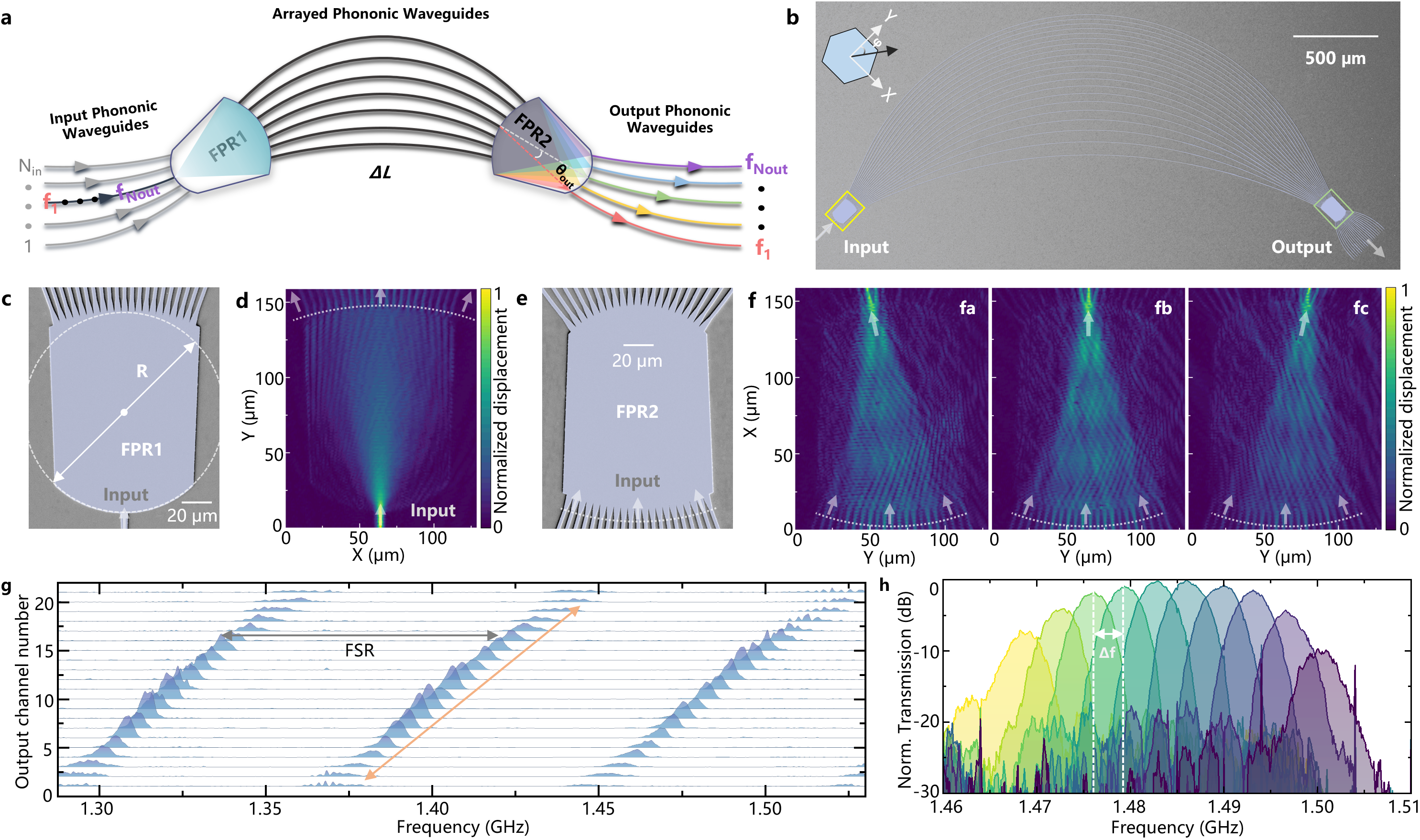}
\caption{\textbf{Acoustic arrayed waveguide grating (AAWG).} \textbf{a},\textbf{b}, Schematic diagram and SEM image of the AAWG. \textbf{c},\textbf{d}, SEM image and corresponding normalized acoustic amplitude distribution in the input free propagation region (FPR) of the AAWG. \textbf{e},\textbf{f}, SEM image and normalized acoustic amplitude distribution at three frequencies in the output FPR: 1.416~GHz ($f_a$), 1.406~GHz ($f_b$), and 1.396~GHz ($f_c$). Acoustic waves of different frequencies converge at different output ports. \textbf{g}, Spectral responses of the 21 AAWG output ports, exhibiting a free spectral range of 81~MHz. \textbf{h}, Zoom-in view of the spectral responses showing a channel spacing ($\Delta f$) of 3.8 MHz between adjacent ports.}
\label{Fig4}
\end{figure*}

\begin{figure*}[t]
\includegraphics[width=1\textwidth]{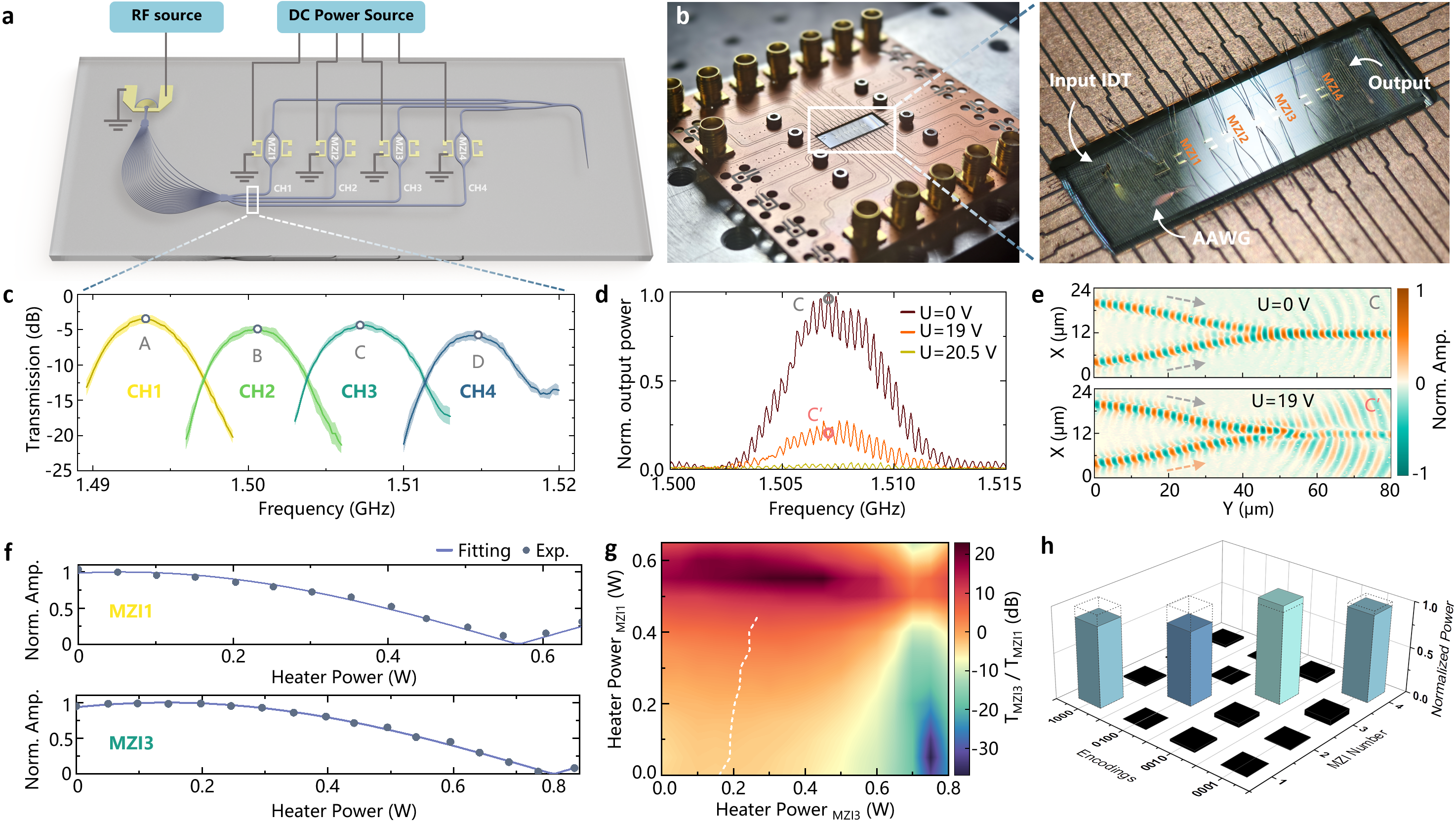}
\caption{\textbf{Reconfigurable acoustic frequency synthesizer (RAFS).} \textbf{a}, Schematic diagram of the RAFS, comprising an AAWG, four thermo-acoustically tunable acoustic MZIs, and a four-to-one acoustic combiner. The input radio frequency signal is converted to acoustic signals through the input IDT and then output from four different ports of the AAWG. The amplitude of the signal from each port is controlled by subsequent MZIs and then combined into a single output waveguide through an acoustic combiner. \textbf{b}, Optical image of the wire-bonded RAFS device. Inset: magnified view of the device. \textbf{c}, Experimentally measured transmission spectra at the four output ports of the AAWG. \textbf{d}, Normalized transmission of MZI3 with different driving voltages on its heater. \textbf{e}, Acoustic interference amplitude distribution at the output port of MZI3 corresponding to points $C$ and $C'$ in \textbf{d}. At point $C'$ (19 V), destructive interference causes acoustic energy to leak into the substrate. \textbf{f}, Normalized output amplitude of MZI1 and MZI3 as a function of heater power. \textbf{g}, Transmission ratio between MZI3 and MZI1 with simultaneous thermal tuning of their heaters. The white dashed line shows the evolution of the maximum output power position of MZI3 under different heater powers of MZI1. \textbf{h}, Demonstration of selective single channel activation by independent MZI control.}
\label{Fig5}
\end{figure*}

\smallskip{}
\noindent \textbf{\large{}Large-scale PnIC}{\large\par}

\noindent To demonstrate the scalability of PnICs, we fabricate an ultra-compact 128-port phonon power splitter, as shown in Fig.~\ref{Fig3}a. The device consists of $2^l-1$ Y-splitters cascaded in a binary-tree network of depth $l=7$ to uniformly distribute the input acoustic wave to $2^l$ output channels. Figure~\ref{Fig3}b illustrates a detailed measured field amplitude for acoustic wave propagation in a representative $l=4$ subsection (white dashed frame in Fig.~\ref{Fig3}a), achieving a Y-splitter density about $3300\,\mathrm{cm^{-2}}$. The results confirm the even splitting of the acoustic wave at each Y-splitter, with the output acoustic amplitudes from all 128 ports shown in Fig.~\ref{Fig3}c. Statistical analysis of all output ports in Fig.~\ref{Fig3}d reveals a mean output amplitude of $6.5\,\mathrm{pm}$ with a standard deviation of $1\,\mathrm{pm}$, following a normal distribution. Such small amplitude fluctuations result from accumulated uncertainties in the splitting ratios and propagation losses across the $l-$depth network, indicating excellent fabrication uniformity across all 127 Y-splitters. Figure~\ref{Fig3}e further characterizes each Y-splitter, showing a splitting ratio of $1\pm 0.076$ and an insertion loss of $0.5\pm0.37\,\mathrm{dB}$ at $1.5\,\mathrm{GHz}$, which demonstrates excellent broadband uniformity.

To quantify the loss contributions, we analyze the insertion loss for each layer of the device by averaging the insertion loss across the $2^{l}$ ports on the $l$-th ($l\leq7$) layer in Fig.~\ref{Fig3}f, where $l=1$ corresponds to the results in Fig.~\ref{Fig3}e. We numerically calculate the expected insertion loss as a function of $l$, with given acoustic field intensity propagation losses of $0$, $0.8$, and $2.4\,\mathrm{dB/mm}$, respectively. Finally, we extract a loss of $2.4\,\mathrm{dB/mm}$ for the current device. Our model further predicts that reducing this loss by $2/3$ would lower the total insertion loss of the 128-port splitter to only $10\,\mathrm{dB}$. In this case, losses from the Y-splitters would become the dominant loss mechanism rather than from waveguide propagation.

Beyond the massive spatial parallelism in PnICs signal processing demonstrated by this ultra-compact 128-port splitter, the inherent dispersion of the acoustic waveguides also enables unprecedented frequency-domain manipulation. To harness this capability, we design an AAWG for frequency division multiplexing (FDM). As shown in Fig.~\ref{Fig4}a, the AAWG is analogous to its optical counterpart~\cite{Smit1988,Takahashi1990}, which consists of two free propagation regions (FPRs) connected by an array of waveguides with incrementally increasing path lengths. The principle of the device can be explained by a sequence of three transmission matrices: each FPR acts as a phononic lens, generating an approximate Fourier transformation matrix $M_{IJ}$ that maps the $J$ input ports of an FPR to its $I$ output ports. The waveguide array introduces a linear phase gradient through the path length difference $\varDelta L$ between adjacent waveguides, creating a diagonal phase matrix $M_{aw}$ with elements $M_{aw,kk}=e^{-ik\varDelta\phi}$, where $\varDelta\phi(f)=\beta(f) \varDelta L$ and $k$ is the waveguide index. For acoustic signals input from port ($N_{in}$+1)/2 with $N_{in}$ being the total waveguide number of the input port, frequency dependent routing is achieved when the phase increment satisfies the grating equation: $\beta_{s}(f)d\cdot\theta_{out,n}+\varDelta\phi(f)=2m\pi$, where $\beta_{s}(f)$ is the frequency related propagation constant in the FPRs at frequency $f$, $d$ is the spacing between the adjacent array waveguides at the incident face of the FPR2, $\theta_{out,n}$ is the output angle of the $n$-th output waveguide shown in Fig.~\ref{Fig4}a. Under this condition, the composite transformation $M_{2}M_{aw}M_{1}$ routes the input from port ($N_{in}$+1)/2 to output port $n$. This process demultiplexes the acoustic wave signals of different frequencies into separate output ports, thereby realizing FDM.

Figure~\ref{Fig4}b shows our fabricated AAWG with 1 input port and 21 output ports for a proof-of-principle demonstration. The two FPRs shown in Figs.~\ref{Fig4}c and \ref{Fig4}e are connected by 21 arrayed phononic waveguides. The acoustic signals injected into FPR1 spread symmetrically into all arrayed waveguides (Fig.~\ref{Fig4}d), exhibiting an approximately Gaussian amplitude distribution consistent with simulation.  As an inverse process, the Gaussian-like amplitude profile from the waveguide array entering FPR2 is focused into a specific waveguide port (Fig.~\ref{Fig4}f). Furthermore, different input frequencies (e.g., $f_a=1.416$~GHz, $f_b=1.406$~GHz, $f_c=1.396$~GHz) experience distinct phase gradients within the waveguide array, enabling spectral routing to different output ports.

The frequency dependent responses of all 21 output ports are further investigated over a wider range from 1.29\,GHz to 1.53\,GHz in Fig.~\ref{Fig4}g. The data show two key features. First, according to the grating equation, the transmission spectrum for each output port exhibits periodic peaks with a free spectral range of $FSR=v_\mathrm{g}/\Delta L$. The experimentally measured FSR is about 81~MHz, consistent with the theoretical value of 80~MHz. Second, the output frequency increases linearly with the port index as expected (the yellow line), with enlarged normalized transmission spectra shown in Fig.~\ref{Fig4}h. The measured adjacent channel spacing is $\Delta f=3.8\,\mathrm{MHz}$, consistent with the theoretical prediction of $\Delta f=v_\mathrm{g}/N_{out}\Delta L$. The adjacent channel isolation exceeds $10\,\mathrm{dB}$, and increasing to about 20\,dB for the next nearest neighbor channels. This high frequency resolution corresponds to a quality factor of $400$, which is mainly limited by the designed parameters and can be further improved by at least one order of magnitude through further design optimization of the PnICs.

\smallskip{}
\noindent \textbf{\large{}Reconfigurable acoustic frequency synthesizer }{\large\par}

\noindent Building upon our demonstrated control over spatial and frequency degrees of freedom, we integrated these capabilities into a programmable acoustic frequency synthesizer. The system architecture, as depicted in Fig.~\ref{Fig5}a, integrates an AAWG for frequency demultiplexing, a thermoacoustic MZI modulator array for amplitude modulation, and a multiport combiner for path control. Figure~\ref{Fig5}b shows the fabricated and packaged device. In operation, microwave signals with different frequencies are fed into the AAWG by IDT. The AAGW spatially separates the frequency components into four distinct channels (Fig.~\ref{Fig5}c), showing an average insertion loss of 4.55~dB. Each channel is then routed to an individual MZI modulator (MZI1-MZI4), whose transmission is controlled by integrated heaters~\cite{Shao2022,Xu2025}. By applying DC biased voltages to these heaters, we tuned the relative phase between each acoustic MZI arms, enabling complete suppression of transmission over a large bandwidth, as shown in Fig.~\ref{Fig5}d. This suppression is achieved through destructive interference of MZI, which leaks the acoustic waves into the substrate (Fig.~\ref{Fig5}e).


Figure~\ref{Fig5}f demonstrates the precise individual frequency channel control by the heater power-dependent transmission of MZI1 and MZI3 at frequencies corresponding to the points $A$ and $C$ in Fig.~\ref{Fig5}c. The blue curves represent the fitting results of the output amplitude accoring to the function $\sqrt{(1+\mathrm{cos}(\alpha P_{H}+\varphi_{0}))/2}$, where $P_{H}$ represents the applied heater power, $\alpha$ quantifies the thermoacoustic coefficient, and $\varphi_{0}$ accounts for the initial phase offset. In Fig.~\ref{Fig5}g,  we simultaneously apply different voltages to the heaters of MZI1 and MZI3 to demonstrate the capability to parallel control each channel. The measured transmission ratios between the two channels $T_{MZI3}/T_{MZI1}$ show a wide rage of tunability from -36.78~dB to 22.86~dB. Furthermore, as the heater power applied to MZI1 increases from 0 to 0.45~W, the power required to achieve the maximum output of MZI3 shifts from 0.16~W to 0.27~W, as the white dashed line shown in Fig.~\ref{Fig5}g. Although this shift indicates thermal crosstalk between the two channels, a common challenge in integrated thermoacoustic devices, the system still maintains sufficient channel isolation for practical applications. As a conceptual demonstration of the frequency synthesizer, we manipulate all four channel outputs at frequencies corresponding to points $A$, $B$, $C$, and $D$ in Fig.~\ref{Fig5}c. We selectively set one channel to ``on" while switching the others ``off". The results (Fig.~\ref{Fig5}h) demonstrate average on/off ratios of 18~dB, 29~dB, 15~dB, and 24~dB for the four cases, respectively. Therefore, these results unambiguously demonstrate a reconfigurable acoustic frequency synthesizer. This system, which leverages the MHz level resolution of the AAWG, reveals the potential of programmable PnICs for synthesizing arbitrary waveforms and performing advanced frequency-domain operations.

\smallskip{}
\noindent \textbf{\large{}Discussion}{\large\par}

\noindent From developing basic building blocks to demonstrating large-scale programmable PnICs with spatial, amplitude, and frequency parallelism at the system level, our work transforms phononic technology from component-level applications to a complete information processing paradigm as a complementary to their electronic and photonic counterparts. Benefiting from tight mode confinement and low propagation loss, PnICs enable ultra-compact, dispersion engineered devices that can surpass the capabilities of conventional approaches. Looking ahead, even larger scale PnICs could directly convert RF signals into acoustic waves to perform signal processing through physical wave propagation, interference, and filtering, potentially enabling phononic AI accelerators with massively parallel, ultra-low power information processing capabilities.

Beyond signal processing applications, PnICs offer a unique platform for hybrid integration. In this envisioned architecture (termed ``Zhengfu"), a monolithic chip could combine phononic, electronic, and photonic circuits~\cite{Xu2022}. In this system, co-propagating acoustic and optical modes would enable efficient photon-phonon conversion via Brillouin interactions~\cite{Yang2024a,Rodrigues2025,Kaixuan2025}, while piezoelectric coupling would provide direct phonon-electron transduction~\cite{Gustafsson2014,Satzinger2018}. This positions phononic circuits as an interface between the photonic communication and electronic processing domains. The quantum potential of PnICs is particularly compelling at cryogenic temperatures, where acoustic losses are dramatically reduced. Strong coupling between superconducting qubits and phononic waveguides has been theoretically predicted, opening a path toward hybrid quantum processing units. In such systems, phonons could serve as coherent mediators between optical photons and qubits, enabling new architectures for distributed quantum computation while providing inherent isolation from electromagnetic noise. This vision positions PnICs not merely as an alternative technology, but as an essential component in the future landscape of quantum information processing.

%

\noindent \textbf{\large{}Acknowledgment}{\large\par}
\noindent This work was funded by the National Natural Science Foundation of China (Grants No. 92265210, 92165209, 92365301, 123B2068, 12104441, 12061131011, 12474498, 92565301, 11925404), the Innovation Program for Quantum Science and Technology (Grant Nos. 2021ZD0300203 and 2024ZD0301500). This work is also supported by the Fundamental Research Funds for the Central Universities, USTC Research Funds of the Double First-Class Initiative, and Beijing National Laboratory for Condensed Matter Physics (2024BNLCMPKF007). The numerical calculations is performed with the Supercomputing Center of USTC. This work was partially carried out at the USTC Center for Micro and Nanoscale Research and Fabrication.

\clearpage{}

\onecolumngrid
\renewcommand{\thefigure}{S\arabic{figure}}
\setcounter{figure}{0}
\renewcommand{\thepage}{S\arabic{page}}
\setcounter{page}{1}
\renewcommand{\theequation}{S.\arabic{equation}}
\setcounter{equation}{0}
\setcounter{section}{0}

\end{document}